\documentclass[aps,a4paper,superscriptaddress,showpacs,showkeys,10pt]{revtex4}
\usepackage{graphicx}
\usepackage{subfigure}
\usepackage{latexsym}   
\usepackage{amsmath}    

\begin{document}

\title{The extended gaussian ensemble and metastabilities in the Blume-Capel model}
\author{Rafael B. Frigori}
\email{frigori@utfpr.edu.br} 
\affiliation{Universidade Tecnol\'ogica Federal do Paran\'a, \\
             Rua XV de Novembro 2191, CEP 85902-040 Toledo, PR, Brazil.}
\author{Leandro G. Rizzi}
\email{lerizzi@usp.br} 
\author{Nelson A. Alves}
\email{alves@ffclrp.usp.br} 
\affiliation{Departamento de F\'{\i}sica e Matem\'{a}tica, FFCLRP, \\
             Universidade de S\~ao Paulo,
             Avenida Bandeirantes, 3900 \\ 
             14040-901, Ribeir\~ao Preto, SP, Brazil.}

\begin{abstract}
   The Blume-Capel model with infinite-range interactions presents analytical solutions in both canonical and microcanonical ensembles and therefore, its phase diagram is known in both ensembles.
   This model exhibits nonequivalent solutions and the microcanonical thermodynamical features present 
peculiar behaviors like nonconcave entropy, negative specific heat, and a jump in the thermodynamical temperature.
   Examples of nonequivalent ensembles are in general related to systems with long-range interactions that undergo canonical first-order phase transitions.
   Recently, the extended gaussian ensemble (EGE) solution was obtained for this model.
   The gaussian ensemble and its extended version can be considered as a regularization of the microcanonical ensemble.
   They are known to play the role of an interpolating ensemble between the microcanonical and the canonical ones.
    Here, we explicitly show how the microcanonical energy equilibrium states  related to the metastable and unstable canonical solutions for the Blume-Capel model are recovered from EGE, which presents a concave  ``extended'' entropy as a function of energy.
    
\end{abstract}


\maketitle

\section{Introduction} 
  Thermodynamical properties of physical systems are described in Statistical Mechanics by the ensemble theory.
  Its development rests on the existence of both additivity and thermodynamic limit, which are essential ingredients to describe critical phenomena.
  The existence of a thermodynamic limit is related to the extensivity property 
which is required in order to obtain meaningful physical results in that limit \cite{Touchette305}.
  Large enough systems with short-range interactions present the additivity property for
extensive quantities like energy and entropy. 
  Additivity means that when we consider the energy, or any other extensive quantity of a system, and divide 
the system into two or more macroscopic parts, the total energy is still the sum of the energies of those parts.
  From theoretical backgrounds, the canonical and grand-canonical ensembles approximate the 
microcanonical ensemble in the limit of infinitely large number of particles, where 
surface effects and fluctuations can be disregarded with respect to the bulk mean values
\cite{Gross_book,Gross_report}.
  There is definitely more consensus around the microcanonical ensemble as the most fundamental one.
  Thus, the canonical and grand-canonical ensembles are taken as approximations for the microcanonical ensemble whenever the full equivalence does not hold.
  In cases where full consistency of statistical ensembles holds for systems that undergo phase transitions, it is found that finite-size scaling relations still place the microcanonical approach
as the fundamental one \cite{kastner_JSP2001}.

  There are many examples of systems whose microcanonical equilibrium properties contrast to the ones obtained in the canonical ensemble. 
  Differences in the equilibrium properties have been described analytically for systems with long-range interactions \cite{Ruffo_PRL2001,Ruffo_PRL2005,Ruffo_JSP2005,Touchette_A2007},
  which are consequences of the nonconcavity of the entropy as a function of energy 
 \cite{Ruffo_PRL2001,Bouchet_JSP118,kastner_A2007}.
  The nonequivalence appears for those systems when the canonical ensemble exhibits a 
first-order phase transition and this has been demonstrated by means of the large deviation theory \cite{TurkingtoJSP101}.
  Such nonequivalence results in uncommon features like temperature discontinuity and negative specific 
heat in the microcanonical ensemble  
\cite{Ruffo_PRL2001,kastner_A2007,TurkingtoJSP101,Ruffo_EPJB2008,Ruffo_Lecture2002,Ellis_A335}.

    In this paper, we will be dealing with a less popular ensemble, the extended gaussian 
ensemble (EGE) \cite{Vives_PRE68}, which is constructed upon the gaussian ensemble \cite{Hether_JLTP66,Hether_PRD35,Hethe_87,Challa_PRL60,Challa_PRA38}.
They are known to present a smooth interpolation  between their limiting behaviors, corresponding to the microcanonical and canonical ensembles.
    Recently, the EGE solution was obtained for the Blume-Capel (BC) model with infinite-range interactions \cite{Frigori2010}. 
    Our numerical calculations illustrate this limiting behavior in a model with a tricritical point.
    This tricritical point does not coincide numerically in the two standard ensembles.
    Here, we explicitly show how the microcanonical energy equilibrium states related to the metastable and unstable canonical solutions for the Blume-Capel model are recovered from EGE, which presents a concave ``extended'' entropy.  
    Moreover, the discontinuous microcanonical solution is also investigated and the metastable and 
unstable lines are also obtained from EGE.

    The content of our presentation is as follows.
    The EGE analytical solution for the BC model is reviewed in the next section.
    Analytical and numerical results for thermodynamical quantities characterizing this ensemble 
are presented in section 3. 
    In section 4 we summarize the results of this methodology in identifying the region of nonequivalence of ensembles only from the EGE point of view.

\section{Extended gaussian ensemble and the Blume-Capel model}

     The EGE is defined by the condition $\gamma \neq 0$ and probability density
\begin{equation}
  P_{\gamma,\alpha}(E) = \frac{\rho(E)\, e^{-\alpha E - \gamma (E- U)^2}}{Z_{\gamma}(U,\alpha)} ,
             \label{eq:prob}
\end{equation}
where $Z_{\gamma}(U,\alpha)$ stands for the normalization constant, which is the corresponding partition function in 
EGE \cite{Vives_PRE68,Touchette_JDP119,Touchette_E73}.
   As usual, $\rho(E)$ is the density of states.
   Here, $\gamma$ and $\alpha$ are parameters, and $U$ stands for the mean energy, which is $\alpha$ and $\gamma$ dependent, $U= U(\alpha,\gamma)$.   
   The extended gaussian ensemble is a particular case in a class of ensembles defined by
 functions $g(E)$ \cite{Touchette_JDP119,Touchette_E73}, where the quadratic form
$g(E)=\gamma(E-U)^2$ is a convenient choice.
   The mean energy $U$ can be obtained from the extended Gibbs's measure,
\begin{equation}
    U = \int E\, P_{\gamma,\alpha}(E) dE   .         \label{eq:U_self} 
\end{equation}

   This ensemble has two important limits. 
   The canonical results correspond to  $\gamma \rightarrow 0$ and in this 
case the parameter $\alpha$ identifies the inverse thermodynamical temperature, $\alpha = 1/k_B T$.
  On the other hand, the limit $\gamma \rightarrow \infty$ corresponds to the microcanonical case.
  This can be seen as 
$ {\rm lim}_{\gamma \rightarrow \infty} \sqrt{\pi/\gamma} Z_{\gamma}(U,\alpha) $
through the use of the Dirac's delta sequence in the gaussian form \cite{Lukkarinen_99}.  
  The finite $\gamma$ case presents an intermediary thermal description between the known limiting
ensembles. 
   All thermodynamical quantities can be defined in this ensemble as well.
   The  extended thermodynamical potential is analogously defined to the canonical ensemble,
 $ \Phi_{\gamma}(U,\alpha) = - {\rm ln}\, Z_{\gamma}(U,\alpha)$.
   From here, the derivative at fixed value $\gamma$ gives the mean energy,
\begin{equation}
    \left(\frac{\partial \Phi_{\gamma}}{\partial \alpha} \right)_{\gamma}  = U  , \label{eq:U}
\end{equation}
which parallels that of the usual canonical approach.
   The extended heat capacity is \cite{Vives_PRE68,Challa_PRL60}
$ C_{\gamma} = - \alpha^2 ({\partial U}/{\partial \alpha})_{\gamma}$.

  Let us now consider the Blume-Capel model \cite{blume_1966,capel_all}, a particular case of the Blume-Emery-Griffiths model \cite{beg_1971}.
  Here, we work with its infinite-range interaction version,
\begin{equation}
  H(S)= \Delta\sum_{i=1}^{N}S_{i}^{2} -\frac{J}{2N}\left(\sum_{i=1}^{N}S_{i}\right)^{2} , 
                                                                            \label{eq:H(S)}
\end{equation}
where $S_i = \{0,\pm 1 \}$.
  The couplings  $J> 0$ and $\Delta$ are the exchange and crystal-field interactions,
respectively.
  This is a well studied model, which presents a rich phase diagram in the $(\Delta/J, T/J)$ plane.
  It exhibits a first-order phase transition line, tricritical point, and a second-order phase transition line.  
  The critical properties of the BC model have been studied analytically in both the microcanonical \cite{Ruffo_PRL2001} and canonical ensembles \cite{beg_1971,beg_1974}.
  The BC model has become an outstanding model for what concerns the nonequivalence of ensembles.
  The canonical and microcanonical ensembles do not yield the same phase diagram for the first-order 
critical line \cite{Ruffo_PRL2001}.
  The canonical tricritical point occurs at $(\Delta/J, T/J) \simeq (0.46209812, 1/3)$,
which gives origin to the canonical first-order phase transition line for larger values of $\Delta/J$
(and lower values of $T/J$). 
  The microcanonical solution identifies the tricritical point at  $(\Delta/J, T/J) \simeq (0.46240788,0.33034383)$.
  
  The EGE analytical solution \cite{Frigori2010} is obtained from the extended partition function by means of Hubbard-Stratonovich transformations.
   To this end, we introduce the order parameters magnetization $M=\sum_{i=1}^{N} S_i= N_+ - N_{-}$, 
and the quadrupole moment $Q= \sum_{i=1}^{N}  S_i^2 = N_+ + N_{-} $, 
where $N_+$ and  $N_{-}$ are, respectively, the number of sites with up and down spins.
   Also, let $N_0$ be the total number of zero spins, then
$N = N_+ + N_{-} + N_0 $ is the total number of spins in the system.
  Therefore, the extended partition function reads
\begin{equation}
Z_{\gamma}(U,\alpha)  =   \sum_{N_{0}=0}^{N}\sum_{N_{+}=0}^{N-N_{0}}
               \frac{N!}{N_{0}!N_{+}!N_{-}!}  e^{-\alpha (\Delta Q -\frac{J}{2N} M^2)  -  
                   \gamma(\Delta Q-U-\frac{J}{2N} M^2)^2}  .     \label{IV_9}  
\end{equation}
   Here, it is convenient to work with the intensive quantities $q= Q/N$ and $ m=M/N$.
   Let us also define $K=J/2\Delta$ and $\varepsilon = U/\Delta N$ the energy per site, following the authors 
in \cite{Ruffo_PRL2001}.
  
   First of all, we recall the EGE solution for $\gamma \rightarrow \infty$, which yields the known microcanonical ensemble results.
  The extended partition function must be well-behaved in this limit. 
  Thus, it is required in equation (\ref{IV_9}) that
   $\varepsilon = q - K m^2 $, which becomes a
constraint equation for the mean energy $\varepsilon$ as $\gamma \rightarrow \infty$.
   The thermodynamic limit  $N \rightarrow \infty$ 
can be evaluated and one obtains the thermodynamical potential per site
\begin{equation}
 \varphi(\varepsilon,\alpha,m)  =  \varepsilon \alpha\Delta          
  + \left[q\ln\left(\frac{\sqrt{q^{2}-m^{2}}}{2(1-q)}\right)
  +   \frac{m}{2}\ln\left(\frac{q+m}{q-m}\right)+\ln\left(1-q\right)\right] ,   \label{f_NonEQ}
\end{equation}
as a function of mean energy per site $\varepsilon$ and, parameters $\alpha$ and mean magnetization $m$.
   The term inside the brackets is the correct microcanonical entropy $- s_{\rm micro}(\varepsilon, m)$ obtained in \cite{Ruffo_PRL2001} as a function of $\varepsilon$ and $m$.
   The entropy is an even function of $m$ and a nonconcave function of independent variables 
$m$ and $\varepsilon$.
   The nonconcave behavior is the necessary condition for metastability and instability in the canonical states 
\cite{Touchette_A365} and gives origin to the nonequivalence of ensembles (microcanonical and canonical).
   The consequences on the thermodynamical quantities specific heat and specific susceptibility have been analyzed in \cite{Ruffo_PRL2001,Frigori2010}.

    The solution for finite $\gamma$ describes up to what extension one recovers the microcanonical stable states.
    The full equivalence of these results with the stable states of the microcanonical ensemble 
is achieved for {\it finite} $\gamma$ only for $\Delta/J$ between the canonical 
and the microcanonical tricritical points. 
    For couplings $\Delta/J$ larger than the microcanonical tricritical point, i.e., 
on the microcanonical first-order transition line, one needs $\gamma \rightarrow \infty $
for such full recovery of the microcanonical states \cite{Frigori2010}.

    The finite $\gamma$ solution for the thermodynamical potential per site reads
\begin{eqnarray}
\varphi_{\gamma}(\varepsilon,\alpha,m,q) & = & q \ln\left(\frac{\sqrt{q^2-m^2}}{2(1-q)}\right)
         +\frac{m}{2} \ln\left(\frac{q+m}{q-m}\right) + \ln(1-q)                         \notag  \\ 
   & &  +\, \gamma \Delta^2 (q- Km^2-\varepsilon)^2 + \alpha \Delta(q-Km^2)  .   \label{phi_gamma}
\end{eqnarray} 
    
  The EGE analytical solution poses a clear way of studying the formation of stable energy-dependent microcanonical states as the parameter $\gamma$ increases. 
  Furthermore, it is possible to draw the EGE solutions to observe the discontinuous microcanonical transition.
   This will be illustrated in the next section.

\section{Results and discussion}

  The lack of ensemble equivalence is a consequence of a nonconcave entropy.
  As discussed in \cite{Touchette_A365}, the possible stable states, labeled by $\varepsilon$, are the ones 
defined by the microcanonical entropy $s_{\rm micro}(\varepsilon)$.
  If those values of $\varepsilon$ are not global minima of the canonical thermodynamical potential per site
as we vary the temperature, then we do not find such correspondence between these states in both ensembles.
  Local minima and local maxima in the canonical thermodynamical potential per site are, respectively, metastable and unstable states.

\begin{figure}[t]
\begin{center}
\begin{minipage}[t]{3in}
\includegraphics[width=3in]{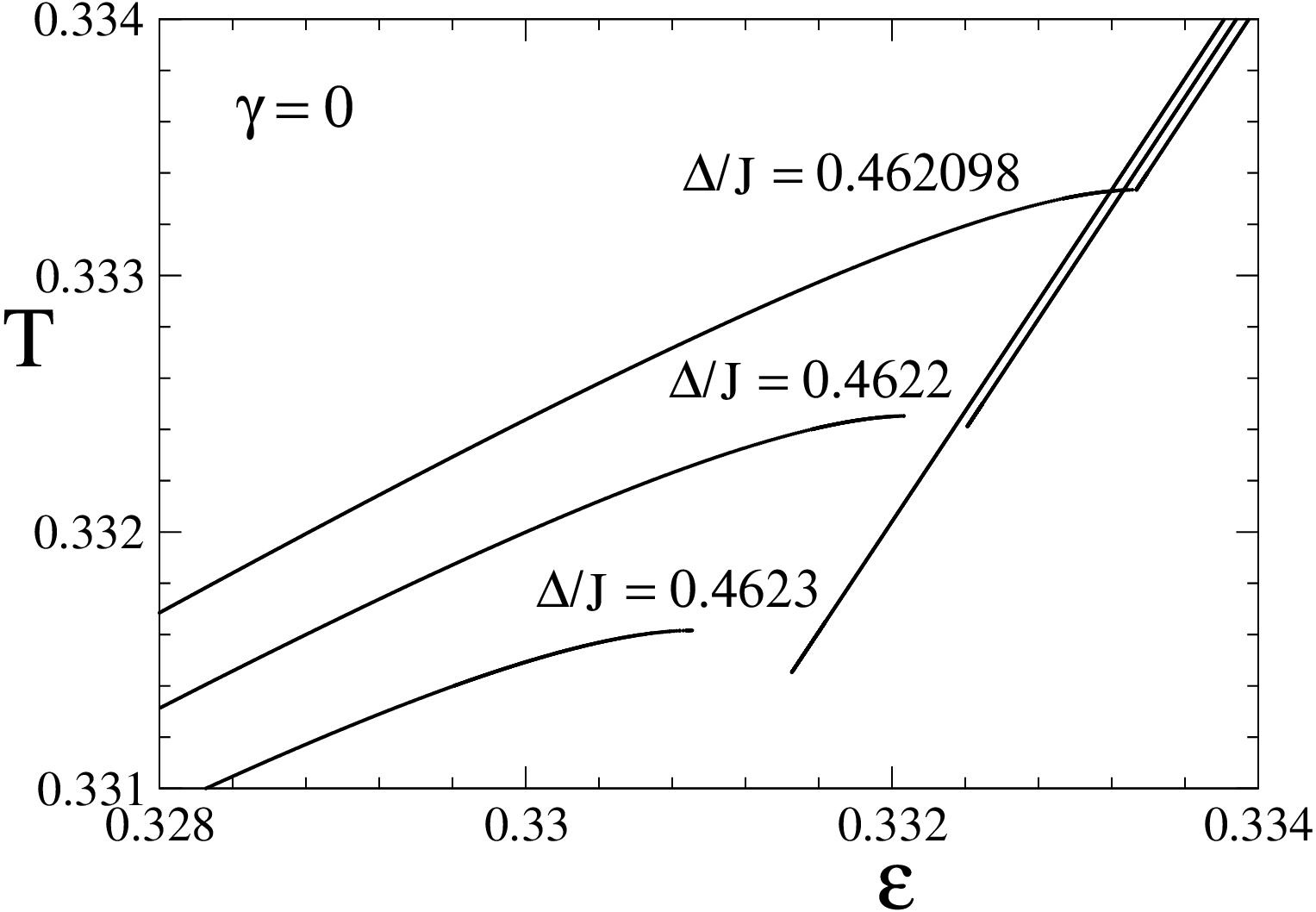}
\caption{EGE temperatures obtained in the canonical limit for some values of $\Delta/J$. $\Delta/J=0.462098$ is in both canonical and microcanonical second-order phase transition regions.
 $\Delta/J=0.4622$ and $0.4623$ are in the canonical first-order phase transition region but in the   
 microcanonical second-order one.}
\label{fig:1}
\end{minipage}
\hspace{.2in}
\begin{minipage}[t]{3in}
\includegraphics[width=3in]{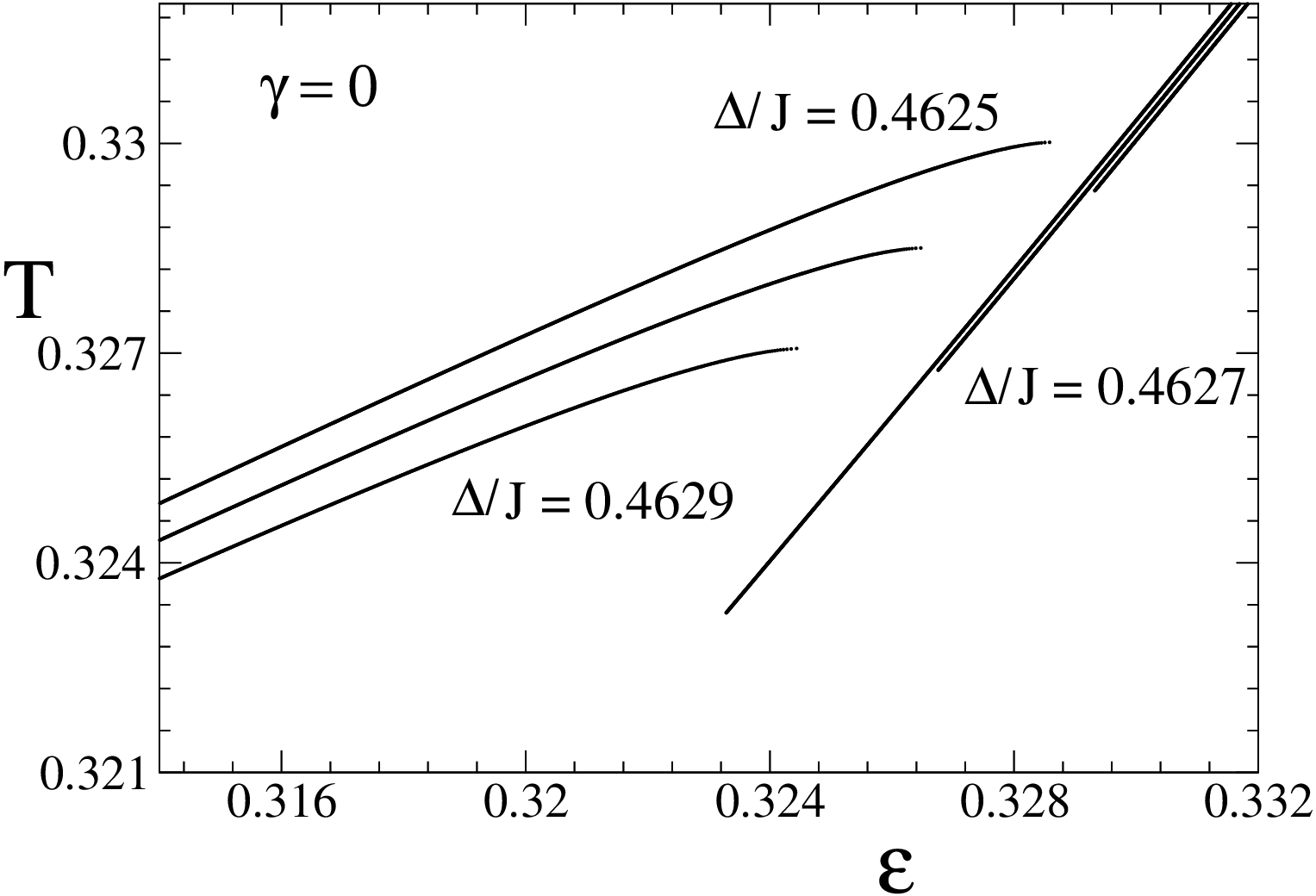}
\caption{EGE temperatures obtained in the canonical limit for some values of $\Delta/J$. The values of $\Delta/J$ are in both canonical and microcanonical first-order 
  phase transition regions.}
\label{fig:2}
\end{minipage}
\end{center}
\end{figure}

  To see how the nonconcave entropy for the BC model induces 
this nonequivalence of ensembles, we initially evaluate the caloric curve
$T$ {\it versus} $\varepsilon$, in the canonical limit.
  To this end, we calculate the equilibrium solutions of $ \varphi_{\gamma}(\varepsilon,\alpha,m,q)$
for $\gamma \rightarrow 0$.
  The equilibrium solutions are easily obtained in this limit and are shown in figures 1 and 2.
  Figure 1 shows caloric curves for different values of $\Delta/J$.
  Notice that the point $\Delta/J = 0.462098$, although close to the canonical tricritical point, 
is still in the second-order phase transition region according to both canonical and microcanonical predictions.
  The points $\Delta/J = 0.4622$ and $0.4623$, are in the canonical first-order phase transition region, but they are in the second-order region from the microcanonical point of view.  
  Figure 2 shows caloric curves for $\Delta/J$ in a region where both ensembles present first-order 
phase transition lines.
  Notice that we have not drawn in those figures horizontal lines corresponding to the canonical Maxwell construction.
  Such constructions can be appreciated in \cite{Ruffo_PRL2001} for some values of $\Delta/J$.
  The missed stable microcanonical states are the ones related to the gaps in $\varepsilon$ in both figures.
  We anticipate that the presentation of those stable microcanonical states is left to figure 4, 
while in figure 5 we included all the equilibrium solutions.
  
  The EGE equilibrium states for finite $\gamma$ can be obtained by
solving the extremum equations for $ \varphi_{\gamma}(\varepsilon,\alpha,m,q)$.
  We call attention to the fact that the microcanonical constraint for the specific quantities is not 
enforced here, the variables $\varepsilon,m$ and $q$ are treated as independent variables in this approach.
  The linear term in $\gamma$, equation (\ref{phi_gamma}), can be seen as a constrained equation leading to the microcanonical ensemble only for $\gamma \rightarrow \infty$.
   
   As done in \cite{Frigori2010}, the desired solutions must satisfy equation (\ref{eq:U}),
$ \partial \varphi_{\gamma}/ {\partial \alpha} = \varepsilon \Delta$,
and the extremum conditions
${\partial \varphi_{\gamma}}/ {\partial m} = 0 $, and 
${\partial \varphi_{\gamma}}/ {\partial q} = 0 $,
for a fixed $\varepsilon$. 
    The stability criterion in the canonical ensemble for the BC model is nicely discussed in
 \cite{Ruffo_Lecture2002}.
    We must observe that the solutions for the three equations above are not stable for all $\gamma$.
    Since the analytical expression $\varphi_{\gamma}$ comes from a saddle-point approximation, one needs to study the stability of those EGE solutions as a function of $m$ and $q$.
    To this end, the determinant of the Hessian matrix,
\[
 d(m,q)
    =\det\left(\begin{array}{cc}
\frac{\partial^{2}\varphi_{\gamma}}{\partial m^{2}} & \frac{\partial^{2}\varphi_{\gamma}}{\partial m\partial q}\\
\frac{\partial^{2}\varphi_{\gamma}}{\partial q\partial m} & \frac{\partial^{2}\varphi_{\gamma}}{\partial q^{2}}
\end{array}\right)  ,
\]
is analyzed in the $T$ versus $\varepsilon$ plane as a function of $\gamma$.
    This amounts to exploring which points $\{m,q\}$ minimize $\varphi_{\gamma}$ for fixed $T$ and
$\varepsilon$, and satisfy the condition $d(m,q) \geq 0$.

\begin{figure}[t]
\begin{center}
\includegraphics[width=3in]{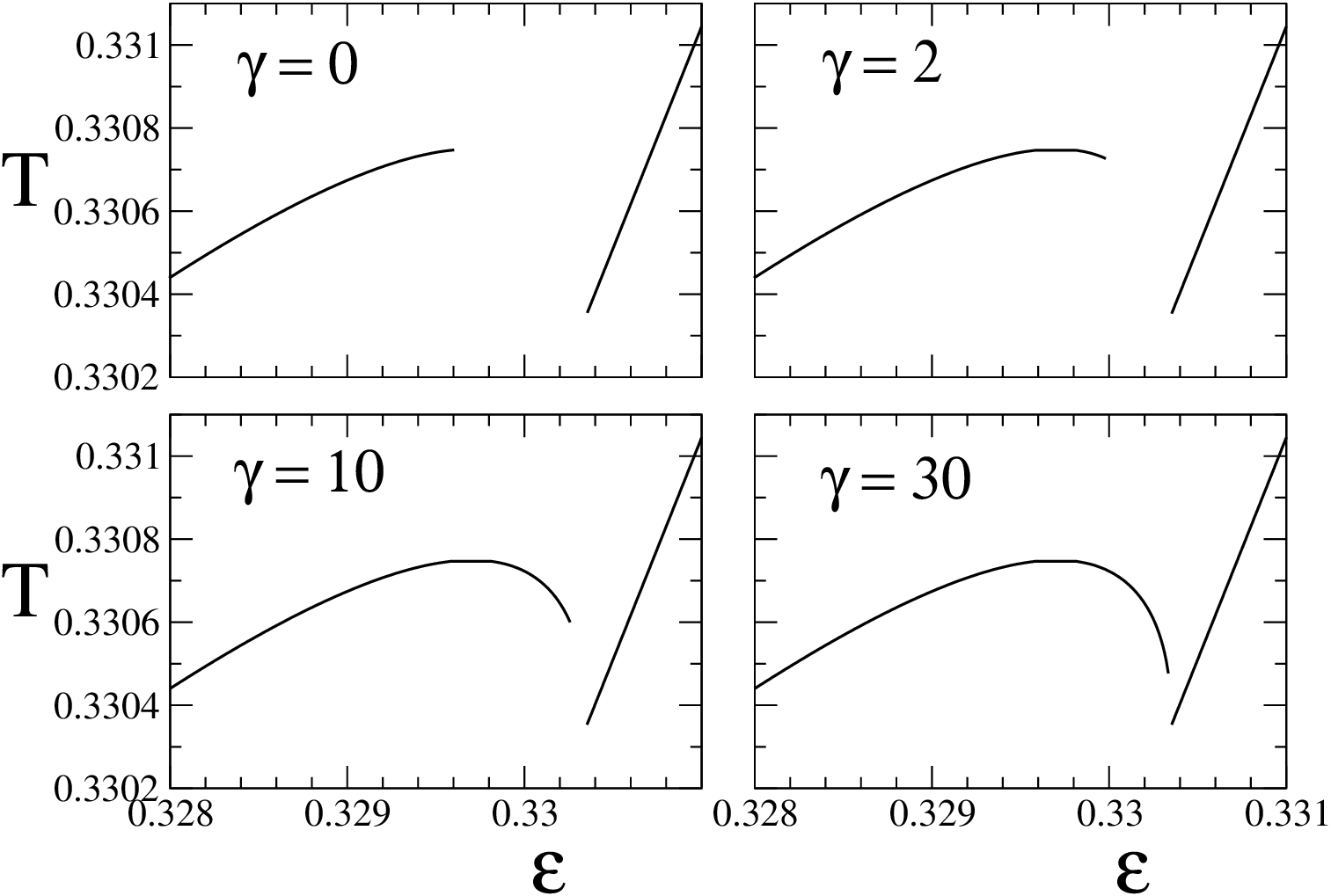}\hspace{2pc}%
\begin{minipage}[b]{4in}
  \caption{\label{fig:3}EGE temperatures for some values of $\gamma$ with $\Delta/J = 0.462407$.
   A finite $\gamma \simeq 4950$ is necessary to recover all microcanonical states.}
  \vspace{.3in}
\end{minipage}
\end{center}
\end{figure}


   Figure 3 collects our results of stable points for different values of $\gamma$.
   That behavior is illustrated for $\Delta/J=0.462407$,
a coupling close to the microcanonical tricritical point and on the canonical first-order critical line, but
still on the second-order critical line according to the microcanonical calculations.
   This figure shows that, as $\gamma$ increases, one recovers the microcanonical solution.
   It can be shown that for sufficiently large $\gamma$, $s_{\gamma}(\varepsilon)$
becomes entirely concave and continuous on $\varepsilon$ \cite{Touchette_E73,Costeniuc07},
\begin{equation}
   \frac{\partial^2 s_{\gamma}}{\partial \varepsilon^2} < 0  .     \label{gammaP}     
\end{equation}
  In other words, the addition of the term in $\gamma$ to the thermodynamical potential 
changes the energy range where its nonconcavity is observed.
  This energy range is reduced as $\gamma$ increases.
  Equation (\ref{gammaP}) can be used to find the necessary condition on $\gamma$ to obtain
the microcanonical solution,
\begin{equation}
   \gamma >  \frac{-1}{2\Delta  T^2(\varepsilon) c(\varepsilon)} .
\end{equation}
 Here, $c(\varepsilon)$ is the specific heat, a negative quantity for energy ranges in figure 3 such that
 $dT/d{\varepsilon} < 0$.
    Figure 3 shows that, for the coupling $\Delta/J=0.462407$, a finite value of $\gamma$ 
is able to recover all stable microcanonical states.
    Actually, $\gamma \simeq 4950$ is the minimum value necessary for this end.
    To obtain the corresponding stable microcanonical states for $\Delta/J=0.4622$ and $0.4623$ in 
figure 1, one needs a rather small value for $\gamma$.
    It can be shown \cite{Frigori2010} that $\gamma \sim 30$ is sufficient for such couplings $\Delta/J$.
    The numerical solutions are presented in figure 4.

\begin{figure}[t]
\begin{center}
\begin{minipage}[t]{3in}
\includegraphics[width=3in]{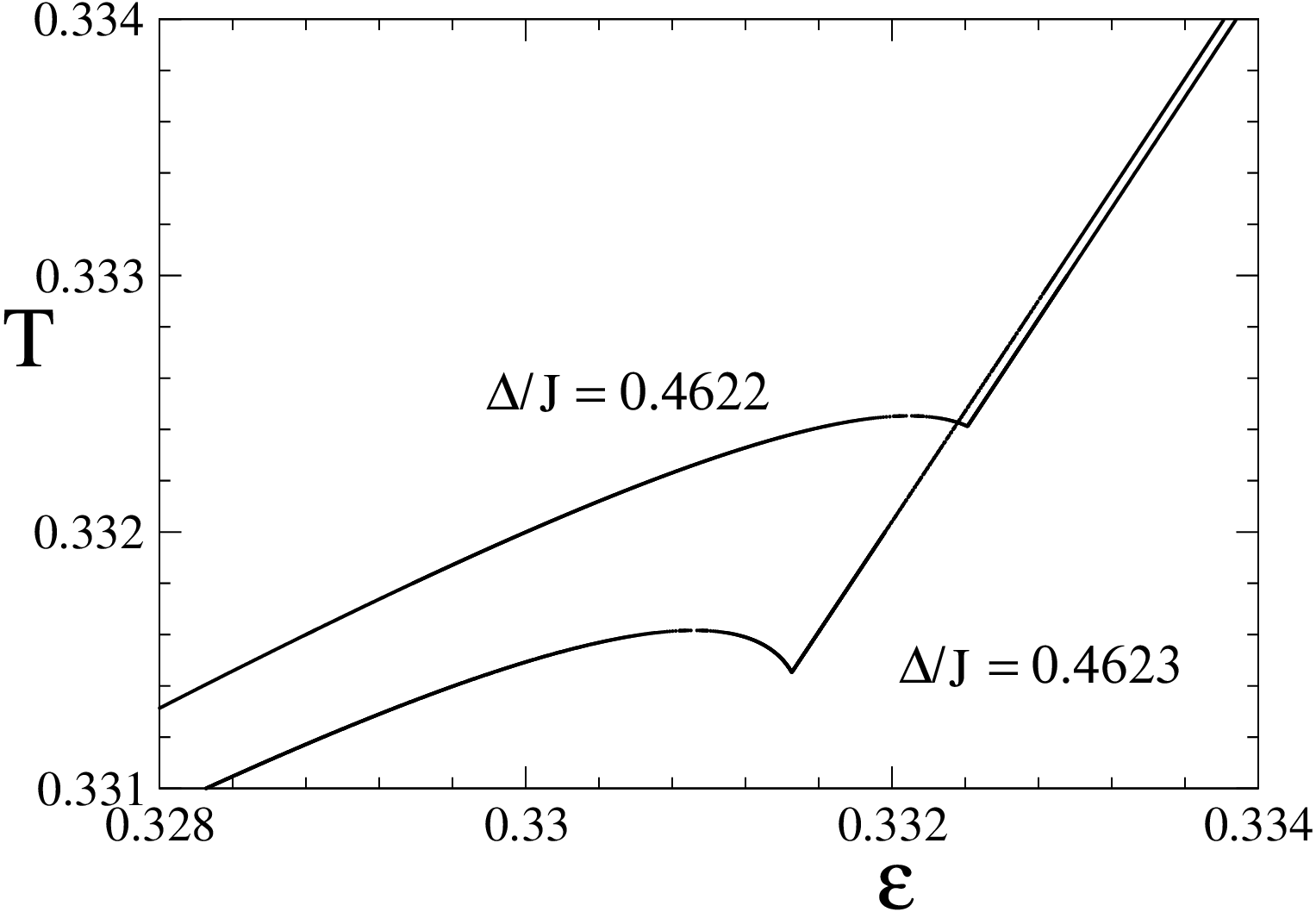}
\caption{EGE temperatures obtained in the microcanonical limit for $\Delta/J=0.4622$ and $0.4623$. The couplings $\Delta/J$ are in the microcanonical second-order phase transition region.}
\label{fig:4}
\end{minipage}
\hspace{.2in}
\begin{minipage}[t]{3in}
\includegraphics[width=3in]{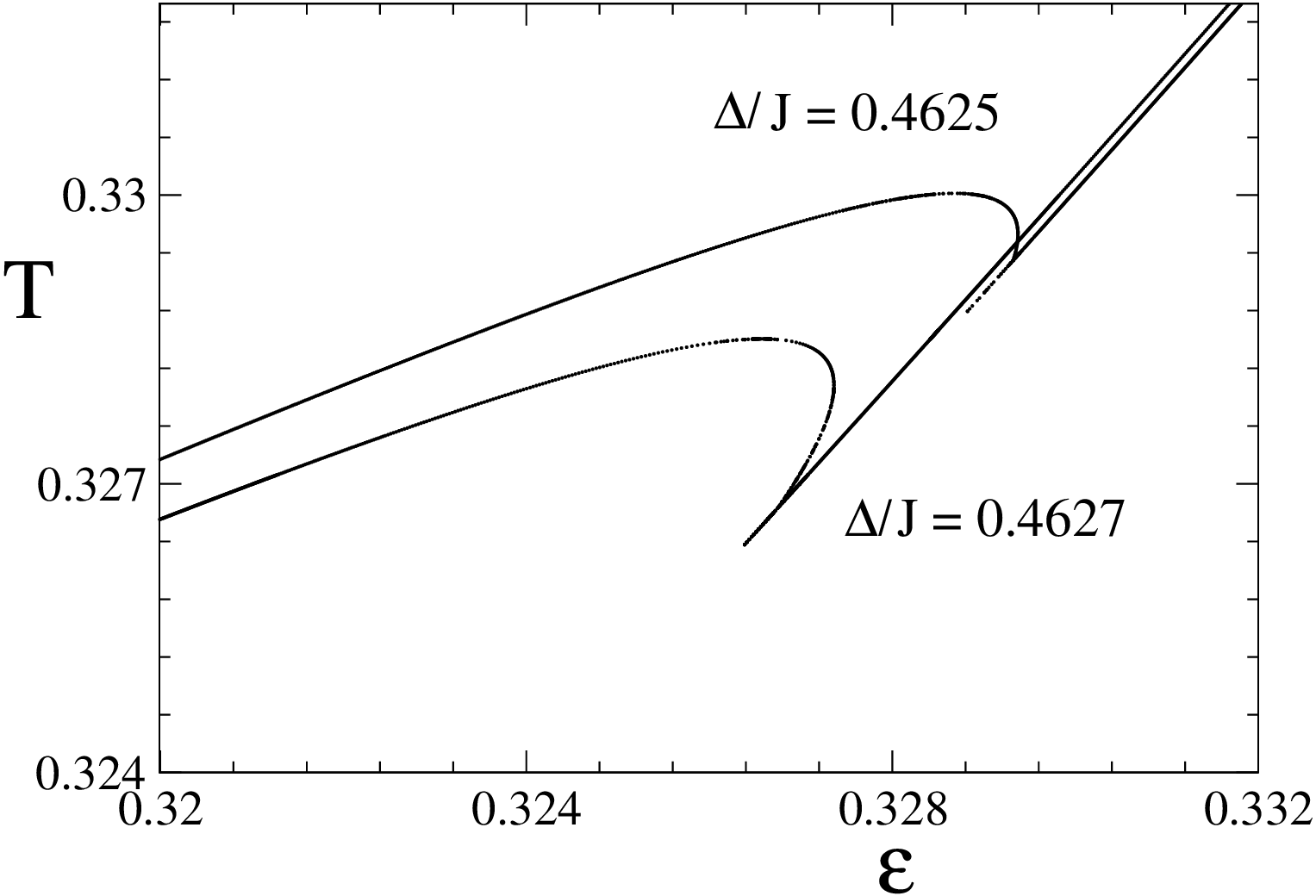}
\caption{EGE temperatures obtained in the microcanonical limit for $\Delta/J=0.4625$ and $0.4627$. The couplings $\Delta/J$ are in the microcanonical first-order phase transition region.}
\label{fig:5}
\end{minipage}
\end{center}
\end{figure}

    For $\Delta/J$ larger than $0.46240788$, the microcanonical tricritical point,
the recovery of all microcanonical states is achieved only for $\gamma \rightarrow \infty $.
    Different of a continuous recovery of microcanonical states, as shown in figures 3 and 4, 
the caloric curves on this microcanonical first-order phase transition line sets a new physical situation.
    Here, one observes jumps in the temperature, the analogous of the latent heat in the canonical ensemble.
    The energy where this transition occurs can be obtained by defining a vertical line corresponding to the Maxwell construction for the microcanonical ensemble \cite{Ruffo_Lecture2002,Chavanis_2006}.
    Moreover, the  microcanonical hysteresis effect is also present.    
    This effect in the thermodynamics of a self-gravitating system has been discussed by Chavanis \cite{Chavanis_2006}.
    Our figure 5 illustrates this situation for $\Delta/J = 0.4625$ and $0.4627$.
    Although for such values of $\Delta/J$ the full recovery of {\it all} microcanonical states is only possible in the limit  $\gamma \rightarrow \infty $, we can still obtain valuable estimates about those states by choosing a very large numerical value for $\gamma$.
    Moreover, we were able to draw the continuation lines that define two branches, the high and low temperature branches, which produce the unstable and metastable lines of microcanonical states.
    The solutions depicted in figure 5 show that $T(\varepsilon)$ is a multivalued function of 
$\varepsilon$, signaling a microcanonical first-order phase transition, in addition to the standard profile of the canonical first-order phase transition when one draws horizontal lines corresponding to the Maxwell's construction.
    
\section{Conclusions}

    The EGE approach leads to analytical expressions for the extended free energy and entropy in a simple way, 
and quantifies the nonequivalence of ensembles.
    The nonconcavity of the microcanonical entropy function means that the system contains several energy-dependent stable states, revealed in the microcanonical ensemble, which have their counterpart in the 
canonical ensemble, but they are unstable or metastable in this latter ensemble.
    By microcanonical entropy we do not refer to the standard Legendre transform of the canonical thermodynamical potential, which always produces a concave function of energy.
  The new term in $\gamma$ added to the canonical free energy turns such unstable or metastable points into equilibrium points in the extended ensemble \cite{Touchette_JDP119,Touchette_E73,Touchette_A365,Touchette_E74}
because it presents a concave ``extended'' entropy.
   Moreover, it is explicitly shown for the BC model that not only the stable states present in the
microcanonical approach can be obtained from EGE but also the unstable and metastable ones.
   This is illustrated in figure 5, which shows the complexity of the microcanonical phase diagram and  
the usefulness of EGE for such study. 
 
%
\section{Acknowledgments}

     The authors acknowledge support from the Brazilian agencies FAPESP, CAPES and CNPq. \\
     


\end{document}